\documentstyle[namedreferences,psfig]{kapproc}

\renewcommand{\cite}[1]{\citeauthor{#1} (\citeyear{#1})}
\renewcommand{\[}{\begin{equation}}
\renewcommand{\]}{\end{equation}}
\renewcommand{\thefootnote}{}

\newcommand{\vc}[1]{\mbox{\bf #1}}
\newcommand{\mx}[1]{\hat{#1}}

\newcommand{\dfrac}[2]{\frac{#1}{#2}}
\newcommand{\tfrac}[2]{{\textstyle\frac{#1}{#2}}}
\newcommand{\ov}{\overline}
\newcommand{\un}{\underline}
\newcommand{\pdv}[2]{\frac{\partial #1}{\partial #2}}
\newcommand{\dpdv}[2]{\pdv{#1}{#2}}

\runningtitle{ANISOTROPY OF TURBULENT CONVECTION} 
\begin{opening}
  \title{THE ANISOTROPY OF LOW PRANDTL NUMBER\\
\bf TURBULENT CONVECTION} 
  \author{Krist\'of Petrovay}
  \institute{Department of Theoretical Physics, University of Oxford\\  
    1 Keble Road, Oxford OX1\,3NP, U.\  K.\\  and \\
    E\"otv\"os University, Department of Astronomy \\
    Budapest, Ludovika t\'er 2, H-1083 Hungary}
\end{opening}

\begin{document}

\footnotetext{\it Geophys.\ Astrophys.\ Fluid Mech.\ 
    \bf 65\rm , 183--201 (1992)}

\begin{abstract}
A model for homogeneous anisotropic incompressible turbulence is proposed. The
model generalizes the GISS model of homogeneous isotropic turbulence; the 
generalization involves the solution of 
the GISS equations along a set of integration paths in wavenumber ($\vc k$-) 
space.  In order to make the problem tractable, these integration paths 
(``cascade lines'') must be chosen in such a way that the behaviour of the 
energy spectral function along different cascade lines should be reasonably 
similar.  In practice this is realized by defining the cascade lines as the 
streamlines of a cascade flow; in the simplest case the source of this flow may 
be identified with the source function of the turbulence.
Owing to the different approximations involved, the resulting energy spectral 
function is not exact but is expected to give good approximative values for the 
\it bulk\/ \rm quantities characterising the turbulent medium, and for the 
measure of the anisotropy itself in particular.

The model is then applied to the case of low Prandtl number thermal 
convection.  The energy spectral function and the bulk quantities 
characterizing the flow are derived for different values of the parameter 
$S=\mbox{Ra}\,\sigma$.  The most important new finding is that unlike the 
anisotropy of the most unstable mode in linear stability analysis
the anisotropy of the turbulence does \it not\/ 
\rm grow indefinitely with increasing $S$ but it rather saturates to a 
relatively moderate finite asymptotic value.  

\end{abstract}

\renewcommand{\thefootnote}{\fnsymbol{footnote}}

\section{INTRODUCTION}
In recent years it has become clear that a better understanding of the 
anisotropy of turbulence in stellar convective zones is important for several 
reasons. It is now widely believed that the 
anisotropy of turbulence plays a key role in governing the transport of angular 
momentum in the convective zones of stars and therefore in determining their 
differential rotation (R\"udiger, 1989).  More accurate knowledge of the 
anisotropy could therefore constrain the theories of differential rotation, 
which at present involve a large number of free parameters. 
Furthermore, it is possible to show (Petrovay, 1990) that the intriguing 
morphological properties of solar and stellar convection found in recent 
numerical simulations (Stein and Nordlund, 1989, Nordlund and Dravins, 1990) 
can be predicted on the basis of sufficiently general convection theories if 
the anisotropy is known.  From its influence on differential rotation 
and on the morphology of convection it follows that the anisotropy of 
turbulence is important in understanding the dynamo mechanism (Brandenburg \it 
et al.\ , \rm 1991, Petrovay, 1991). Finally, some current theories of stellar
convection (e.\  g.\  Xiong, 1980) rely on the assumption that the anisotropy of
turbulence remains moderate throughout the convective layer, while other 
formalisms (Canuto 1989, 1990) include the anisotropy as a free parameter to
be specified.  All this suggests that a method to compute the anisotropy of
turbulence that is simple enough to be applicable in practice while it still
retains the essential physical ingredients of the problem could be very
useful in several astrophysical areas.

It seems to be plausible to treat \it homogeneous\/ \rm anisotropic 
turbulence as a first step.
Despite the fact that homogeneous aniso\-tropic turbulence is ``to some 
extent a non-problem'' (Leslie 1973), in the sense that any process generating 
(aniso\-tropic) turbulence relies on the presence of an inhomogeneity on the large 
scales while the turbulence becomes isotropic on the small scales, it has been 
widely recognized that the study of this subject is of high importance for 
several reasons.  First, it is expected that with increasing $k$ wavenumber 
the turbulence tends to homogeneity faster than to isotropy (Leslie 1973), so 
there exists a wavenumber-regime where the model applies. Secondly, in certain 
flows (e.\ g.\  in grid turbulence sufficiently far from the grid) the 
homogeneous aniso\-tropic approximation may be valid even on the scale of the 
energy-containing eddies.  Thirdly, and perhaps most importantly, the study of 
homogeneous and aniso\-tropic turbulence may be an important intermediate step 
towards the fully general case of inhomogeneous aniso\-tropic turbulence.

As efforts to tackle the problem of anisotropic turbulence on the basis of 
two-point closure models like 
the DIA did not bear much fruit owing to the intimidating complexity of the 
equations, Canuto \it et al.\ \/ \rm (1990) propose that, to begin with, simpler
heuristic models should be used, followed by the construction of more complex 
models (in 
analogy with the historical development of the theory of homogeneous isotropic 
turbulence).  

\begin{figure}
  \centerline{\psfig{figure=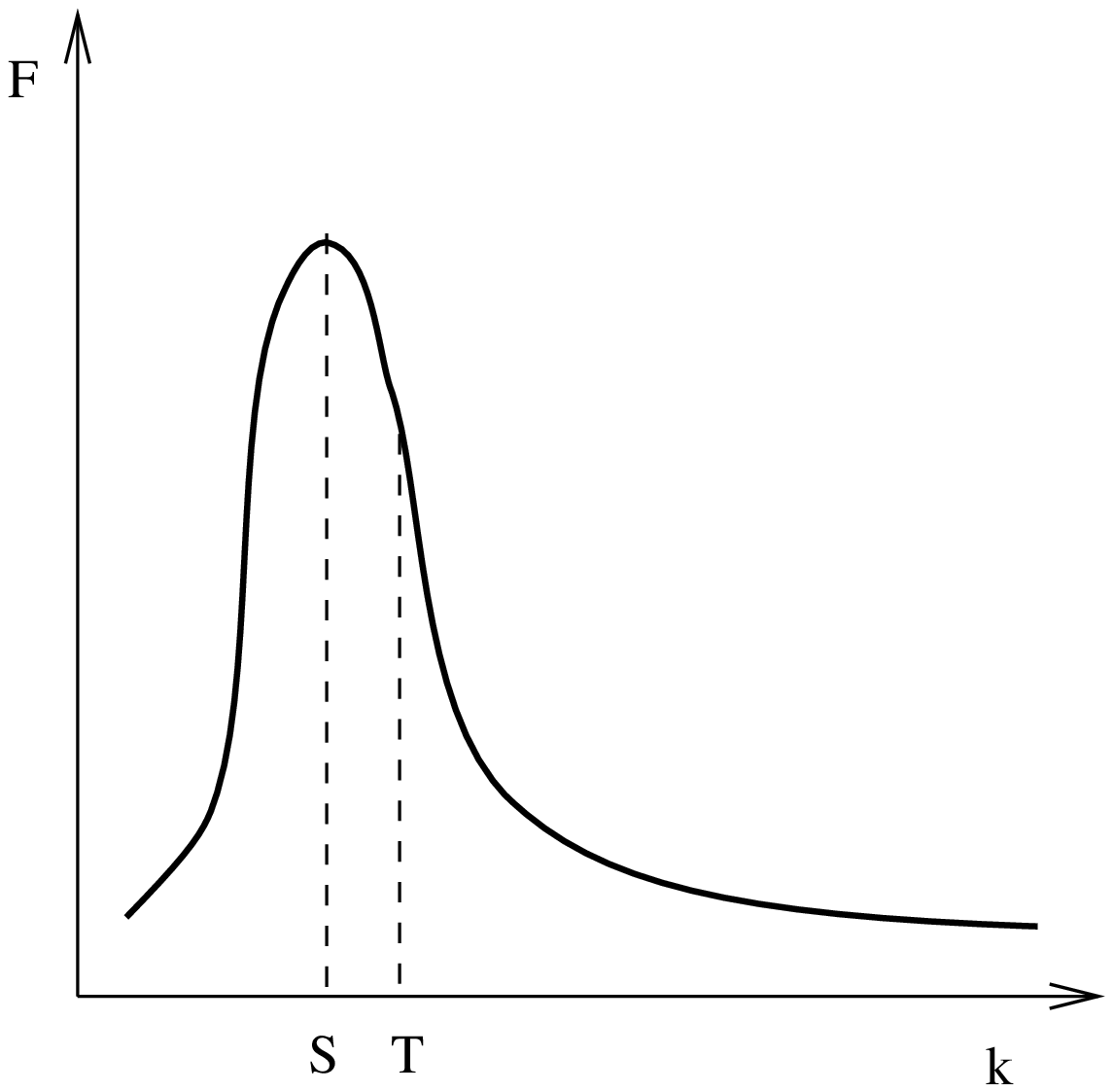,width=6cm,angle=270}
              \psfig{figure=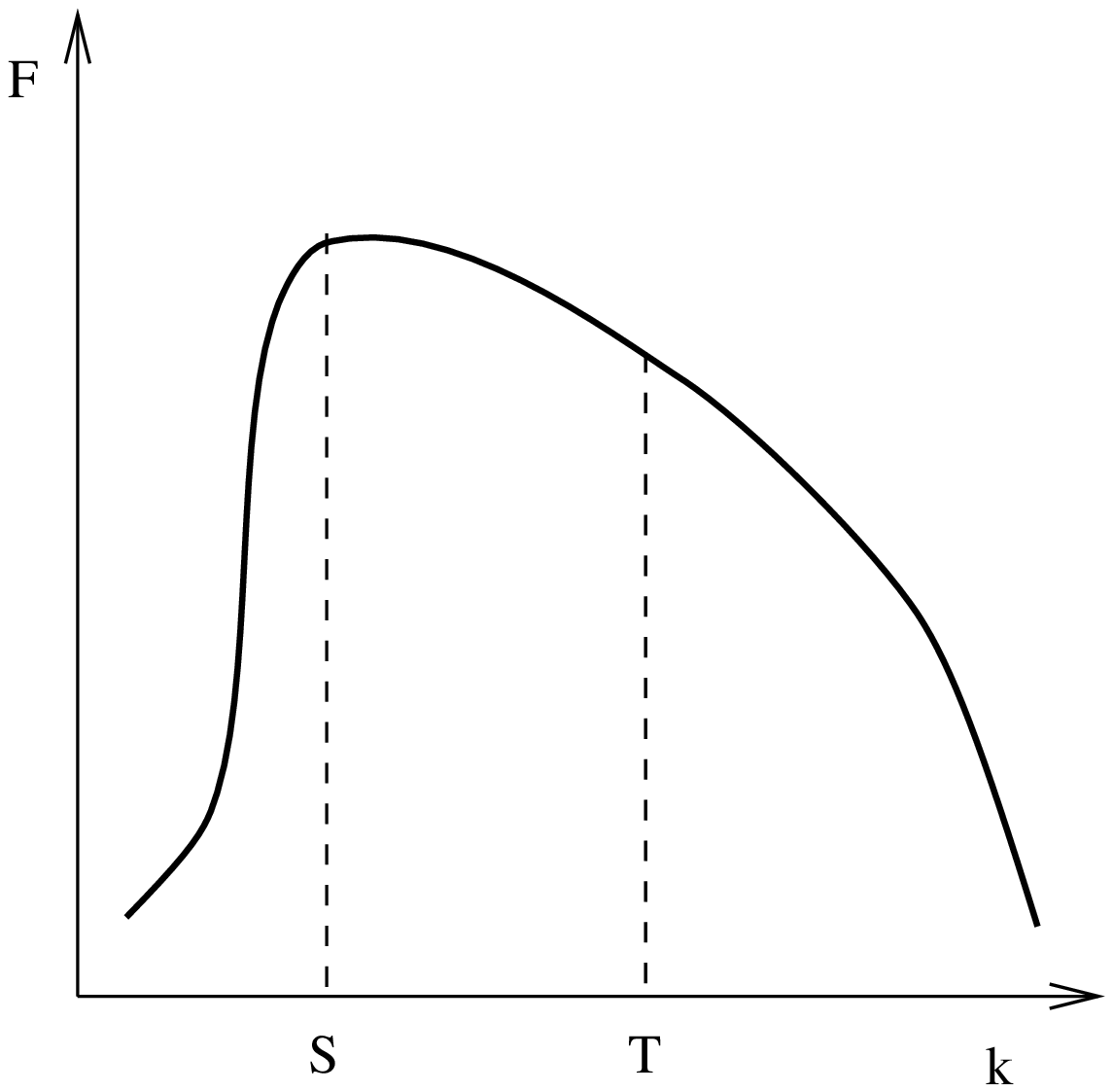,width=6cm,angle=270}}   	
\caption{The anisotropy of the ``typical'' (T) and 
the strongest (S) 
mode may be similar or totally different, depending on how sharply peaked the
spectral function is.}
\end{figure}

The measure of anisotropy is usually characterised by quantities like
\[ x=\frac{l_z^2}{l_x^2+l_y^2}  \]
or
\[ s=\frac{\ov{v_x^2}}{\ov{v_z^2}}  \]
where $l_x$, $l_y$, $l_z$ are ``characteristic scales'' of the turbulence (e.\  
g.\  macroscales or correlation lengths) in different directions, $\vc v$ is the 
turbulent velocity and $z$ is a preferred direction (e.\ g.\  due to gravity). 
In contemporary practice values of these quantities are often ``guessed'' on 
the 
basis of their values for the most unstable mode in linear stability analysis. 
Turbulent values for the diffusivities are sometimes used in the analysis as an
attempt to take the nonlinear interaction of modes into account.  
It is however rather uncertain whether the anisotropy computed in this way
represents the anisotropy of the turbulence correctly, 
not only because nonlinear mode 
interactions may be expected to distort the shape of the spectrum shifting the
maximum, but also because quantities like (1) or (2) are 
defined as \it averages\/ \rm over all wavenumbers, weighted e.\ g.\  by 
\[ \Phi_{ij}(\vc k) = \frac 1{(2\pi)^3}\int \exp(-i\vc k\un\xi) 
   \ov{v_i(\vc r)v_j(\vc r+\un\xi)} \,d^3\xi ,   \]
the velocity covariance spectral function (or its trace $F\equiv \Phi_{ii}$), 
and 
not as values for the single mode corresponding to the maximum of $F(\vc k)$.
The difference is highlighted in Figure 1 (for the sake of clarity in one 
dimension). Depending on how sharply peaked $F(\vc k)$ is, the ``typical'' $\vc 
k$ value may be similar to or totally different from the location of the 
maximum of $F(\vc k)$.

In fact, for the case of low Prandtl number convective turbulence in 
numerical experiments the value of the $s$ parameter defined in 
(2) is found to be very nearly constant at a moderate value of $s\simeq 0.37$ 
for a range of the 
\[ S= \mbox{Ra}\, \sigma = \frac{g\delta\beta D^4}{\chi^2}   \]
parameter covering several orders of magnitude (Chan and Sofia, 1989). 
(Ra is the Rayleigh number, $\sigma$ the Prandtl number, $g$ the 
gravity acceleration, $\delta$ the volume expansion coefficient, $\beta$ the 
superadiabatic temperature gradient, $\chi$ the thermodiffusion 
coefficient\footnote{$\chi=\vert F_m\vert (\rho c_p 
\vert\nabla T\vert)^{-1}$ 
with $\rho$ the density, $c_p$ the specific heat at constant pressure, $T$ the 
temperature and $F_m$ the microscopic (radiative + conductive) energy flux.}, 
$D$ the layer depth, usually identified with the $H_P$ pressure scale height.)
This is in contrast with the anisotropy of the most unstable mode in linear 
stability analysis which grows indefinitely with increasing $S$ (Canuto,
1990). On the other hand, Canuto (1990) has shown that if molecular 
diffusivities are replaced by turbulent ones in the linear analysis, the 
anisotropy of the most unstable mode, though still quite high and slowly 
increasing with $S$, will be seriously reduced.  This was the first theoretical 
result indicating that nonlinear interactions of modes have the potential to 
limit the anisotropy of turbulence.

The present paper proposes a method to 
derive the measure of the anisotropy as well as other bulk quantities in a 
consistent (though approximate) way.  Section 2 contains the general description 
of the method.  This model is
then applied to the astrophysically relevant case of low Prandtl number
convective turbulence in Section 3.  The results, summarized in Section 4, 
agree with the above mentioned numerical simulations.
Section 5 concludes the paper.

\section{THE MODEL}
\subsection{General description of the model}
The model presented below is a generalization of 
the GISS model developed by Canuto, Goldman and Chasnov (1987, hereafter CGC). 
The GISS model is a single-point closure scheme belonging to the class of 
``turbulent viscosity approach'' models; it has been shown to compare well with 
experiments and DIA results (see CGC).

We suppose that $\Phi_{ij}(\vc k)$ is a ``smooth'' function, i.\ e.\  that the 
spectrum of the allowed modes is not discrete.  In cases of astrophysical 
interest, for which this model was primarily developed, this is often a 
reasonable approximation as the studied turbulent regions here form parts of 
much larger systems (e.\ g.\  stars) and so the eigenmodes populate the $\vc 
k$-space densely enough. The assumption may also be reasonably valid in some 
laboratory experiments (e.\ g.\  grid turbulence).  However, for systems with a 
discrete and sparsely populated mode spectrum the present approach cannot be 
applied.

In its principal axis frame the velocity covariance spectral function takes 
the form
\[ \mx\Phi= \left(\begin{array}{ccc} \Phi_1 &0&0\\0&\Phi_2&0\\0&0&\Phi_3 \end{array} 
   \right).    \]
For each $\Phi_i(\vc k)$ component we have an $n_{S,i}(\vc k)$ source function 
or growth rate, supposed to be known from the linear stability analysis. (The 
dissipation rate is also supposed to be built into $n_{S,i}$.)  So we can form 
an $\mx n_S$ source tensor function in such a way that in the principal axis 
frame of $\mx\Phi$ the form of $\mx n_S$ is
\[ \mx n_S= \left(\begin{array}{ccc} n_{S,1} &0&0\\0&n_{S,2}&0\\0&0&n_{S,3} \end{array} 
   \right).  \]
The total energy input per unit time into a given $\vc k$ mode is then 
$\mx n_S(\vc k):\mx\Phi(\vc k)$. ($\mx A:\mx B\equiv A_{ij}B_{ij}$.)

The basic equation of the model expresses the equality of the energy fed into a 
given $\cal V$ volume of the $\vc k$-space with the energy transferred away 
from that volume:
\[ \int_{\cal V} \mx n_S ({\vc k}'):\mx\Phi({\vc k}')\,d^3k'= 
   \int_{\cal V}\,d^3k' \int_{\ov{\cal V}} T({\vc k}', {\vc k}^{''}) 
   \, d^3k^{''} ,  \]
$\ov{\cal V}$ being the complement of $\cal V$.  The $T({\vc k}', 
{\vc k}^{''})$ transfer term is to be determined by the closure.  In the 
turbulent viscosity approach the r.\  h.\  s.\  of equation (6) is taken to be the 
product of the total vorticity of modes in $\cal V$ with the turbulent 
viscosity.

In the case of isotropy equation (7) for a continuous manifold of spheres 
centered on the origin of the coordinate frame uniquely determines $\mx\Phi$, 
once $\mx\alpha\equiv\mx\Phi/F$ is specified.  In the case of anisotropy, 
however, we need several (in general three) such intersecting manifolds of 
surfaces.  Equivalently, one may take a continuous manifold of curves 
originating from a common $P$ starting point and covering the $\vc k$-space. 
This manifold defines a ``curvilinear polar coordinate frame'' with coordinates 
$\kappa$, $\theta_C$, $\phi_C$ ($\kappa$ is the length along each curve 
measured from $P$, the $\theta_C$ and $\phi_C$ ``angles'' effectively index the 
curves). Now in equation (7) we may choose for $\cal V$ a ``curvilinear cone'' 
with angle $\Delta\Omega_C$ centered around the $0<\kappa'<\kappa$ stretch 
of the $(\theta_C, \phi_C)$ curve.  Taking the $\Delta\Omega_C\rightarrow 0$ 
limit and using the notation $k_c^2$ as defined by
\[ d^3k= k_c^2\,d\kappa\,d\Omega_C  \]
we get
\[ \int_0^\kappa \mx n_S (\kappa',\theta_C,\phi_C):\mx\Phi(\kappa',
   \theta_C,\phi_C) k_c^2(\kappa',\theta_C,\phi_C)\,d\kappa'= 
   y(\kappa,\theta_C,\phi_C) \nu_t(\kappa,\theta_C,\phi_C)  \]
where
\[ y(\kappa,\theta_C,\phi_C)= \int_0^\kappa F(\kappa',\theta_C,\phi_C) 
   k^2(\kappa',\theta_C,\phi_C) k_c^2(\kappa',\theta_C,\phi_C)\, 
   d\kappa'    \]
is the total vorticity along the $(0,\kappa)$ stretch of the curve.

According to the turbulent viscosity concept $\nu_t$ can be written in the form 
of an integral of independent contributions.  On a purely dimensional basis
\[ \nu_t(\kappa,\theta_C,\phi_C)= \int_{\ov{\cal V}(\kappa,\theta_C,\phi_C)} 
   \frac{F(\vc k^{''})}{n_c(\vc k^{''})}\,d^3k^{''},   \hfill (11) \] 
$n_c^{-1}$ being a characteristic timescale of the turbulence.  The GISS model 
closure is
\[ \gamma n_c=k^2\nu_t  \]
($\gamma=(2/3\mbox{Ko})^3$ where Ko is the Kolmogorov constant).

Equation (11) shows that in order to solve the system of equations along a 
given path, the value of $F$ must be known along all the other paths.  How 
can we avoid this awkward recursion?

Suppose that our choice of the integration paths is so fortunate that the 
behaviour of $F(\kappa)=F(\kappa,\theta_C,\phi_C)$ is very similar (or even 
identical) for all $(\theta_C, \phi_C)$ paths.  In the case of this 
``fortunate'' choice we will call the paths \it cascade lines. \rm  Equation 
(11) can now be written as
\[ \nu_t(\kappa,\theta_C,\phi_C)= 4\pi \int_\kappa^\infty 
   k_c^2(\kappa^{''},\theta_C,\phi_C) \frac{F(\kappa^{''},\theta_C,\phi_C)}
   {n_c(\kappa^{''},\theta_C,\phi_C)}\,d\kappa^{''} ,   \]
i.\ e.\  the integration over the angles could be replaced by a $4\pi$ factor and 
the recursion has disappeared. The method of finding the cascade lines (i.\ e.\  
such a fortunate choice of paths) will be discussed in the next subsection.

Now the equations (9), (10), (12) and (13) can be transformed into a more  
convenient form, following the treatment in CGC with slight modifications.

Differentiating (9) with respect to $\kappa$ and using the notation $\mx\alpha=
\mx\Phi/F$ we get
\[ (\mx n_S:\mx\alpha)+ 4\pi y n_c^{-1}= \nu_t k^2. \]
From this 
\[ 2\gamma n_c=(\mx n_S:\mx\alpha)+\left[(\mx n_S:\mx\alpha)^2+ 16\pi\gamma 
   y\right]^{1/2}.   \]
Still following CGC we introduce
\[ V=4\pi y+\tfrac 12 \gamma n_c^2 .     \hfill (16)  \]
With this, equation (15) becomes
\[ \gamma n_c= \tfrac 13 (\mx n_S:\mx\alpha)+ \left[\tfrac 19 
   (\mx n_S:\mx\alpha)^2+\tfrac 23 \gamma V\right]^{1/2} .   \]
Now differentiating (16) and using the closure (12)
\[ \pdv V\kappa =\gamma n_c^2\frac 1{k^2} \pdv{k^2}{\kappa} .   \]
(Partial derivatives are taken with $(\theta_C, \phi_C)=$const.\ , i.\ e.\  along 
the cascade lines.)

Were $\mx\alpha(\kappa,\theta_C,\phi_C)$ and $k_c^2(\kappa,\theta_C,\phi_C)$ 
known, equations (16)--(18) would determine $y(\kappa,$ $\theta_C, \phi_C)$, and 
therefore $F(\kappa, \theta_C, \phi_C)$. On the basis of the original integral 
form of the equations the initial condition is
\[ V(\kappa=0)= \frac 1{2\gamma} (\mx n_S:\mx\alpha)^2 .  \]

For $\mx\alpha$ and $k_c^2$, the asymptotic behaviour in the limits 
$\kappa\rightarrow 0$ and $\kappa\rightarrow\infty$ is known (for 
incompressible turbulence). In the intermediate regimes interpolation formulas 
may be used.  These, and the choice of the cascade lines will be discussed in 
the next subsection.

\begin{figure}
  \centerline{\psfig{figure=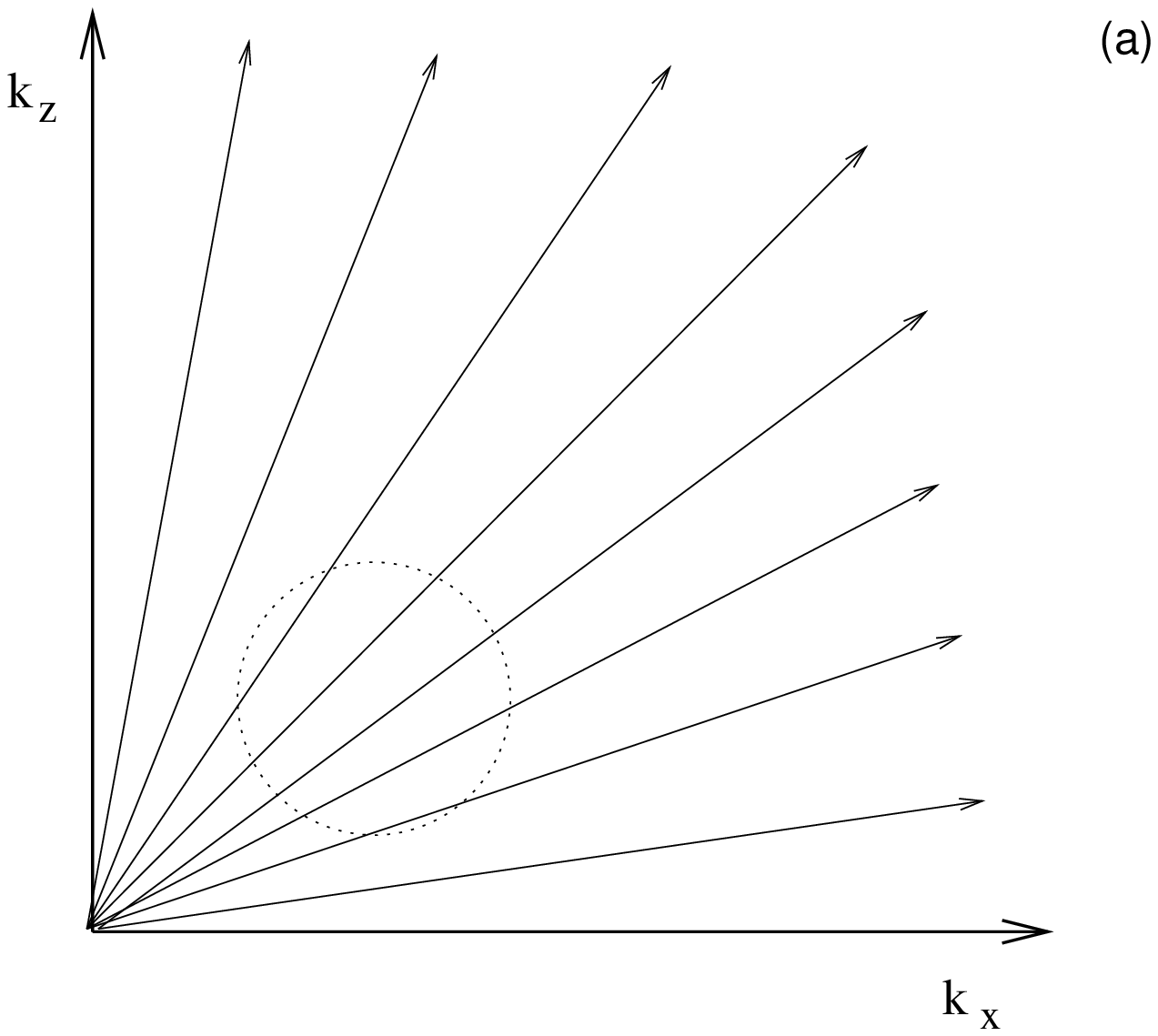,width=6cm,angle=270}
              \psfig{figure=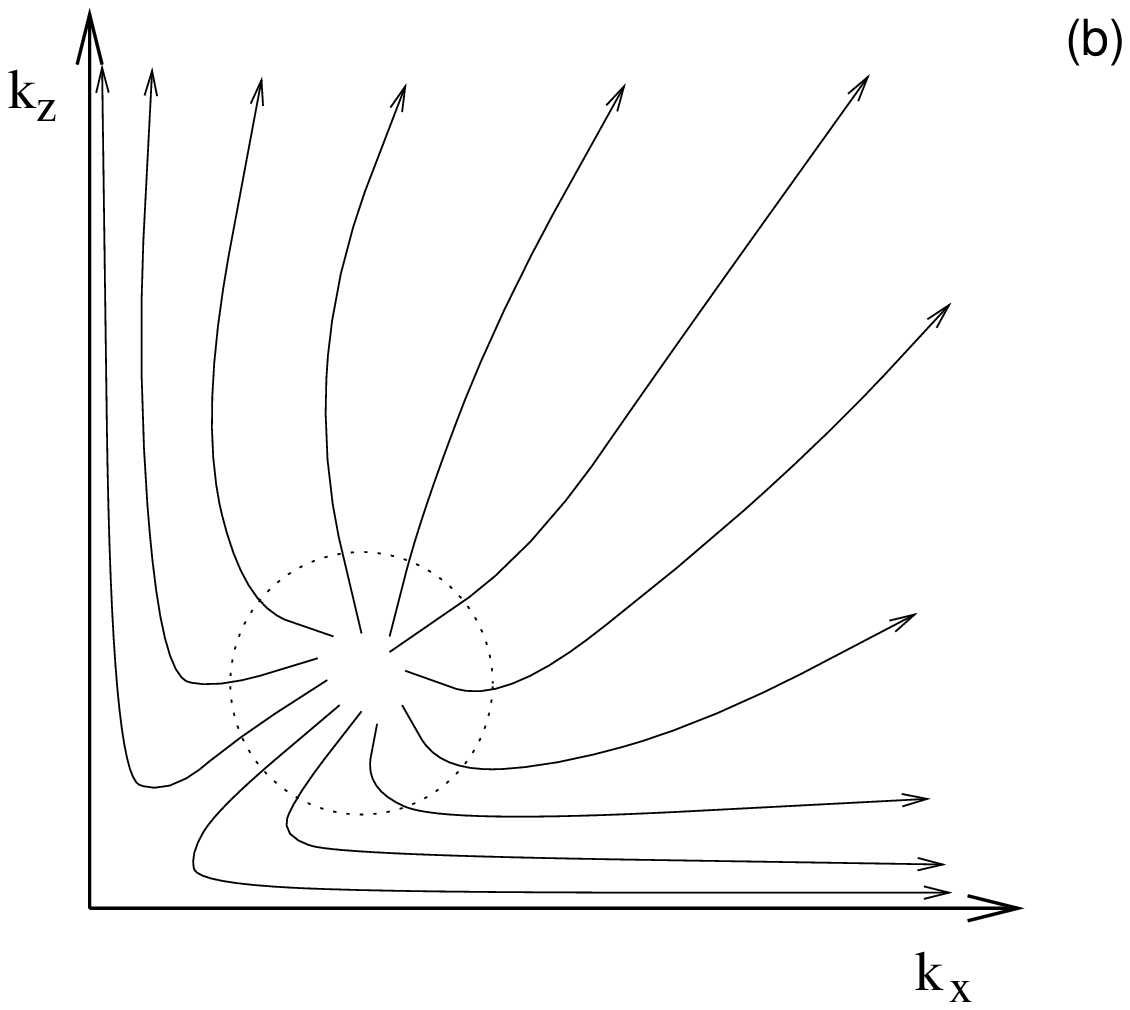,width=6cm,angle=270}}   	
  \caption{Inappropriate (a) and appropriate (b) choices for the cascade 
lines with a source function sharply peaked in the dashed area of $\vc 
k$-space.}
\end{figure}

\subsection{Choice of the cascade lines and quenching}
In the isotropic case the cascade lines may obviously be chosen as straight 
lines starting from the origin (Figure 2a).  The behaviour of $F(\kappa)$ is in 
this case exactly identical for the different cascade lines.  In the 
aniso\-tropic case, however, this choice is not appropriate anymore. If, for 
instance, Tr$(\mx n_S:\mx\alpha)$ has a sharp peak in a certain regime of the 
$\vc k$-space, marked by a dashed outline in Figure 2 (a quite common case in 
physical processes generating turbulence) the energy input along paths crossing 
this regime will greatly exceed the input along other paths, so the $F(\kappa)$ 
values will be very different for different paths at large $\kappa$.  On the 
other hand, one might expect that a choice of the cascade lines that follows 
the general structure of the $\mx n_S(\vc k)$ function defining the problem 
(like the one sketched in Figure 2b) will lead to a much more balanced 
distribution of the energy input between different paths.

These considerations lead us to specify the cascade lines the following way. 
Let the lines be the streamlines of a ${\vc j}_c$ vector field.  As for large 
$k$-s we expect isotropy, and therefore radial cascade lines, we can put 
rot$\,{\vc j}_c=0$ or ${\vc j}_c=\nabla\psi_c$.  The divergence of ${\vc j}_c$ 
is now equated to the source function:
\[ \nabla^2\psi_c=-\mbox{Tr}(\mx n_S).    \]

An equally defensible approach would be to put
\[ \nabla^2\psi_c=-(\mx n_S:\mx\alpha)F   \]
on the basis that it is the total energy input that determines the shape of 
$F(\vc k)$. As however no formula can in general ensure that $F(\kappa)$ will 
be \it exactly\/ \rm identical for every path and equation (22) would involve 
much more numerical work (as using it coupled to the equations (16)--(18) some 
kind of iterative solution would be necessary for $F$), it is more economic to 
use equation (20).  Nevertheless, alternative choices of the cascade lines can 
be used to check the dependence of the results on this particular choice.

Physically we expect that with increasing $k$ wavenumber the turbulence becomes 
more and more isotropic.  This property enables us to find the asymptotic form 
of $k_c^2(\vc k)$ and $\mx\alpha(\vc k)$ for large $k$ values.  

For $k_c^2$ we simply expect $k_c^2\rightarrow k^2$ as $k\rightarrow\infty$; 
while at points in the immediate neighbourhood of $P$ we expect 
$k_c^2\rightarrow k_{c,0}^2=(\vc k-{\vc k}_P)^2$ for truly 3-dimensional 
anisotropy and $k_c^2\rightarrow k_{c,0}^2=\pi k_{x,P}|\vc k-{\vc k}_P|$ for 
axially symmetric turbulence (i.\ e.\  with only one preferred direction).  In the 
intermediate regime we may simply use an arbitrary interpolation formula 
describing the ``quenching'' of the anisotropy towards large $k$-s:
\[ k_c^2=\frac p{1+p} k^2 + \frac 1{1+p} k_{c,0}^2    \]
where
\[ p= \frac{|\vc k-{\vc k}_P|}{Q k_P} ;   \]
$Q$ is a numerical factor which must be order of unity as $k_P$ is basically 
the single important ``length scale'' in $\vc k$-space.

In order to be able to find a similar formula for $\mx\alpha$ we must restrict 
ourselves to incompressible flows for which $\mx\Phi$ has only two nonzero 
independent components, say $\Phi_1=0$ (cf.\  Batchelor, 1953).  In this case 
$\mx\alpha$ will take the form
\[ \mx\alpha= \left(\begin{array}{ccc} 0&0&0\\0& 1-\alpha &0\\0&0&\alpha\end{array} \right) 
     \]
and in the $k\rightarrow\infty$ limit $\alpha\rightarrow 1/2$ while in the 
neighbourhood of $P$ where (in the framework of the present model) all the 
energy in the modes comes directly from the energy input by the instability 
generating the turbulence we have $\alpha=n_{S,3}/(n_{S,2}+n_{S,3})$. Our 
interpolation formula will be
\[ \alpha=0.5 \frac p{1+p} + \frac{n_{S,3}}{(n_{S,2}+n_{S,3})} \frac 1{1+p}.
    \]

The precise functional form of these interpolation formulas is physically not 
too relevant, the sole important parameter that influences the values of the 
bulk properties is the $Q$ quenching parameter.

\subsection{Discussion of the model}
To summarize what was said above, the recipe to compute $F(\vc k)$ for 
homogeneous aniso\-tropic incompressible turbulence goes as follows.

Starting from the $\mx n_S$ source function defining the problem one solves the
\[ \nabla^2\psi_c=-\mbox{Tr}(\mx n_S)   \]
Poisson-equation for the $\psi_c$ cascade potential; physically we expect 
$\psi_c\rightarrow \epsilon/k$ as $k\rightarrow\infty$, so the boundary
condition 
at large $k$ values may be chosen as
\[ \frac 1k \psi_c(k,\theta,\phi) + \pdv{\psi_c(k, \theta,\phi)}k=0 .  \] 
Now taking the gradient of $\psi_c$ one can compute a desired number of cascade 
lines starting from the $P$ maximum site of $\psi_c$. The system
\[ \pdv V\kappa =\gamma n_c^2\frac 1{k^2} \pdv{k^2}{\kappa} .     \]
\[ \gamma n_c= \tfrac 13 (\mx n_S:\mx\alpha)+ \left[\tfrac 19 
   (\mx n_S:\mx\alpha)^2+\tfrac 23 \gamma V\right]^{1/2} .   \]
\[ V=4\pi y+\tfrac 12 \gamma n_c^2 .     \]
\[ F=\frac 1{k^2 k_c^2} \pdv y\kappa   \]
can now be integrated along each cascade line with the initial condition
\[ V(\kappa=0)= \frac 1{2\gamma} (\mx n_S:\mx\alpha)^2 .    \]
The expressions for $k_c^2$ and $\mx\alpha$ are given by equations (22)--(25).

The resulting $F(\vc k)$ can then be used to compute the integrals defining the 
bulk quantities characterising the turbulence, among them the anisotropy 
parameters such as (1) or (2). Let us derive e.\ g.\  the expression for the $s$ 
parameter defined in equation (2). It is a straightforward algebraic exercise 
to compute the components of $\mx\alpha$ in the fixed $k_x k_y k_z$ frame of 
reference (in the incompressible case, using the $k\theta\phi$ polar 
coordinates):
\begin{eqnarray}
   &&\alpha_{xx}=\alpha(\cos^2\theta\cos^2\phi-\sin^2\phi)+\sin^2\phi \nonumber \\
   &&\alpha_{xy}=\alpha\left[(\cos^2\theta+1)\sin\phi\cos\phi\right] 
   -\sin\phi\cos\phi \nonumber \\
   &&\alpha_{xz}=\alpha\sin\theta\cos\theta\cos\phi \\     
   &&\alpha_{yy}=\alpha(\cos^2\theta\sin^2\phi-\cos^2\phi)+\cos^2\phi  \\
   &&\alpha_{yz}=\alpha\sin\theta\cos\theta\sin\phi \nonumber \\
   &&\alpha_{zz}=\alpha\sin^2\theta \nonumber  .
\end{eqnarray}
Using this, for $s$ we get
\[ 2s+1=\frac{\int_0^\infty\,dr \int_0^{2\pi}\,d\phi \int_0^\pi \alpha(r, 
   \theta, \phi) F(r, \theta, \phi) \sin\theta\,d\theta}{\int_0^\infty\,dr 
   \int_0^{2\pi}\,d\phi \int_0^\pi F(r, \theta, \phi) \sin^3\theta\,d\theta}.
    \]

\begin{figure}
   \centerline{\psfig{figure=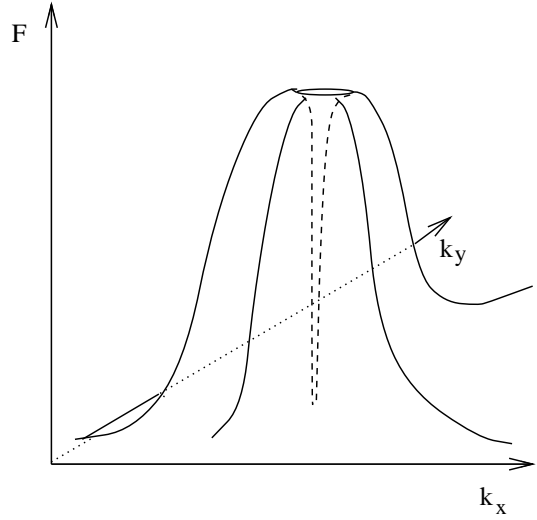,width=7 cm,angle=270}}   	
   \caption{Owing to the assumption of a preferred direction of energy 
transport, a spurious narrow ``hole'' appears near the maximum of $F(\vc k)$ in 
the present model.}
\end{figure}

It is obvious that the resulting $F(\vc k)$ is only a rather crude 
approximation of the real velocity covariance spectral function.  An assumption 
inherent in applying equation (6) to describe the flow is that the energy 
cascade has a preferred direction; this will lead to a spurious ``hole'' in 
$F(\vc k)$ around $P$ where $F(\vc k)$ goes to zero (Figure 3).  This ``hole'' 
is narrow 
enough for its effect on the bulk quantities not to be serious; it basically 
corresponds to the lack of backscatter in the isotropic GISS model.  On the
other hand, the fact that some cascade lines may actually run towards \it 
decreasing\/ \rm $k$ values (cf.\  Figure 2b) means that the aniso\-tropic model 
\it does\/ \rm contain some backscatter---an improvement with respect to the 
isotropic case!

The approximation in equation (12) has the effect that the asymptotically 
isotropic behaviour of $F(\vc k)$ at large $k$ values is not strictly 
guaranteed 
anymore.  As the bulk properties of the turbulence are mainly determined by the 
energy-containing eddies at low $k$ values, this inconsistency is not expected 
to cause serious errors in the bulk quantities. (Yet a quenching of the 
resulting $F(\vc k)$ function with the same $Q$ parameter as in equation  (22) 
may be used to correct this flaw.)

Altogether the method described here seems to be appropriate for a first
approximative derivation of the values of the bulk quantities, most notably the 
anisotropy itself, in a homogeneous, aniso\-tropic, incompressible turbulent 
flow.  

\section{APPLICATION}
For the numerical implementation of the
method we nondimensionalize every quantity by choosing $D =\chi =1$:
\[ \begin{array}{ll}
   \tilde k = kD 
   &{\tilde n}_S = n_S D^2 \chi^{-1} \qquad  \\
   \tilde y =y D^2 \chi^{-2}
   &\tilde\kappa =\kappa D\\
   \tilde F = F D^{-1} \chi^{-2}
   &\tilde V= V D^4\chi^{-2}
\end{array}
\]
etc.  In order to simplify the notation, the tildes will be omitted from here 
on, but we must remember that all the formulae below refer to the 
nondimensional quantities.

The problem at hand is axially symmetric, so in $x,z,\phi$ cylindrical 
coordinates we may restrict ourselves to the $k_xk_z$-plane and ${\mx n}_S$ has 
the form ${n}_{S, xx}={n}_{S, zz}=n_S$; ${n}_{S, \phi\phi}=
{n}_{S, x\phi}={n}_{S, z\phi}={n}_{S, xz}=0$.  The formula for 
$n_S(\vc k)$ was written down in the CGC paper; for low Prandtl numbers it is
\[ n_S(\vc k) = -\frac 12 k^2 + \left[\frac 14 k^4 + S\frac{k_x^2}{k^2}\right].
    \]
As in practice we would like to apply this formula to stratified layers 
(identifying $D=H_P$), a $k_{co}$ cutoff wavenumber is introduced so that for 
values $|k_z| < k_{co}$, $n_S$ is taken to be zero.  This is intended to 
imitate the decrease of the growth rate with increasing vertical eddy size above
a size of $\sim 2 H_P$ in linear stability analysis (Narasimha and Antia 1982; 
see also the Figure 1 in Chan and Serizawa 1991).  In order to check the dependence 
of the results on the choice of $k_{co}$, the computations were performed with 
two different $k_{co}$ values (1.0 and 3.14).

The $\alpha(\vc k)$ and $k^2_c(\vc k)$ functions are given in equations (22) and (25); 
the $Q$ parameter values studied were 0.5, 1, 2, 3 and $\infty$.

The equations of the model were solved on a $32\times 32$ rectangular grid in 
a square-shaped regime of the $k_xk_z$ plane: $0\leq k_x, k_z \leq 
k_{max}=100$.  As the most interesting behavior of $F(\vc k)$ is expected at 
low $k$ values, the grid was not uniformly spaced; instead, new variables
\begin{eqnarray}
   &&\hat k_x=\log (k_x+1) \nonumber \\
   &&\hat k_y=\log (k_y+1) \nonumber \\
   &&\hat \kappa=\log (\kappa+1)\\
   &&\hat k=\log (k+1)  \nonumber 
\end{eqnarray}
were introduced and the grid spacing was chosen to be uniform in these 
variables. In the new variables equations (27)--(31) and (26) take the form
\[ \pdv V{\hat\kappa} =\gamma n_c^2\frac 1{k^2} \pdv{k^2}{\hat\kappa}  \]
\[ \gamma n_c= \tfrac 13 (n_S\alpha)+ \left[\tfrac 19 
   (n_S\alpha)^2+\tfrac 23 \gamma V\right]^{1/2}     \]
\[ V=4\pi y+\tfrac 12 \gamma n_c^2        \]
\[ F=\frac 1{k^2} \frac 1{k_c^2} \frac 1{\exp \hat\kappa}\pdv y{\hat\kappa} 
    \]
\[ V(\hat\kappa=0)= \frac 1{2\gamma} (n_S\alpha)^2     \]
\[ \exp (-2\hat k_x)\pdv{{}^2\psi_c}{\hat k_x^2}+ \exp (-2\hat k_z)
   \pdv{{}^2\psi_c}{\hat k_z^2}  -\exp (-2\hat k_x)\pdv{\psi_c}{\hat k_x}
   -\exp (-2\hat k_z)\pdv{\psi_c}{\hat k_z}= -2 n_S .     
\]
The boundary conditions for the $\psi_c$ cascade potential are
\[ \begin{array}{cl}
   \qquad \dpdv{\psi_c}{\hat k_x}=0  \qquad &\mbox{ at } \hat k_x=0\\
   \qquad \dpdv{\psi_c}{\hat k_z}=0  \qquad &\mbox{ at } \hat k_z=0\\
   \qquad \dfrac{(k_{max}+1)k_{max}}{[(\exp\hat k_x-1)^2+
    (\exp\hat k_z-1)^2]^{1/2}} 
    \psi_c + \dpdv{\psi_c}{\hat k_z}=0  \qquad &\mbox{ at } \hat k_x=100\\
   \qquad \dfrac{(k_{max}+1)k_{max}}{[(\exp\hat k_x-1)^2+
    (\exp\hat k_z-1)^2]^{1/2}} 
    \psi_c + \dpdv{\psi_c}{\hat k_x}=0  \qquad &\mbox{ at } \hat k_z=100 .
\end{array}
\]
The elliptic problem defined by (43) and (44) was solved for $\psi_c$ by a 
7-point multigrid iteration scheme using Numercal Algorithms (NAG) software 
library routines.  After computing the resulting cascade lines, the (38)--(42) 
GISS equations were integrated along each cascade line with the Euler-Cauchy 
method.  The resulting $F$ values were then used to interpolate $F$ on the mesh 
points of a regular rectangular and a polar grid and the bulk quantities 
characterizing the turbulent flow (weighted integrals of $F(\vc k)$) were 
computed.

The whole procedure was performed for many different values of $S$ in the 
regime $10^2$--$10^{17}$ (the range of $S$ in the solar convective zone) and 
with the $k_{co}$ and $Q$ values quoted above. In order to check the dependence 
of the results on the choice of the cascade lines, in some cases alternative 
definitions for $\psi_c$ were also used (as described in Section 2.2).  As 
expected, the results proved to be fairly insensitive to the choice of the 
cascade potential as long as it by and large reflected the general 
geometrical structure of the source function.

\begin{figure}
   \centerline{\psfig{figure=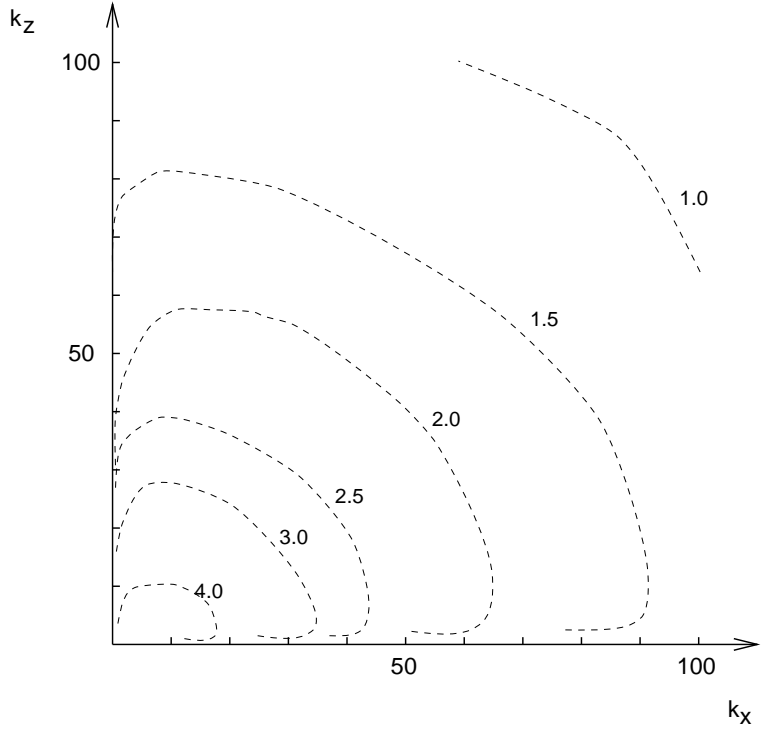,width=10cm,angle=270}}   	
   \caption{Contour lines of the $F(\vc k)$ double energy spectral 
function on the $k_xk_z$ plane for the case $S=10^{10}$, $k_{co}=1$, 
$Q=1$.}
\end{figure}

\begin{figure}
   \centerline{\psfig{figure=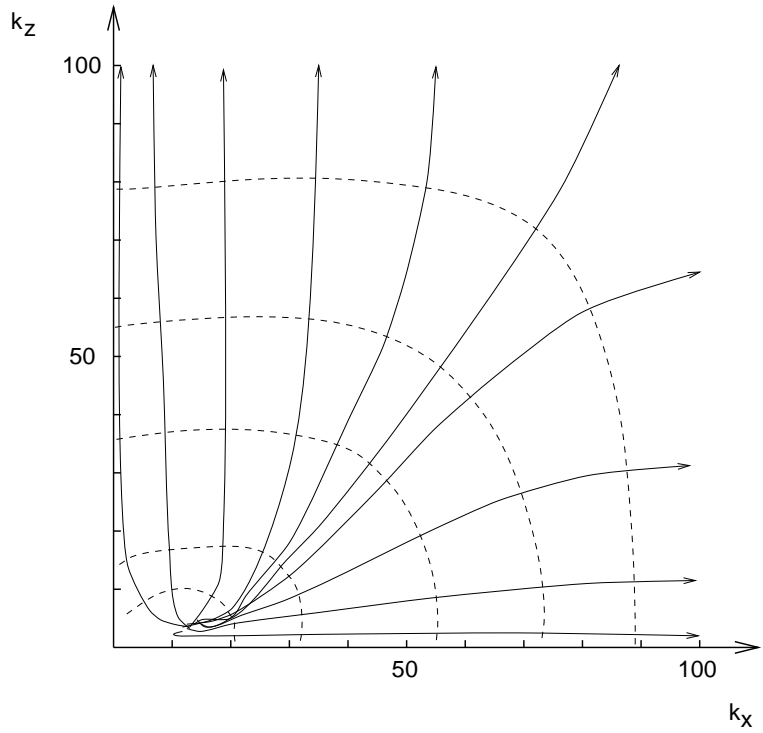,width=10cm,angle=270}}   	
   \caption{Cascade lines and contour lines of the cascade potential 
(dashed) for the same case as in 
Figure 4.}
\end{figure}

\section{RESULTS}
The contour lines of $F(\vc k)$ are shown for an example run in Figure 4.  The 
preference of high $k_x/k_z$ modes over low $k_x/k_z$ modes (i.\ e.\  vertical 
anisotropy) is apparent, particularly at low $k$ values (in the range of the 
energy-containing eddies).  Figure 5 shows the cascade lines for the same 
example. As explained in Section 2.4., the present model does not ensure the 
isotropic behavior of the solution as $k\rightarrow\infty$, but this has little 
effect on the bulk properties.

\begin{figure}
   \centerline{\psfig{figure=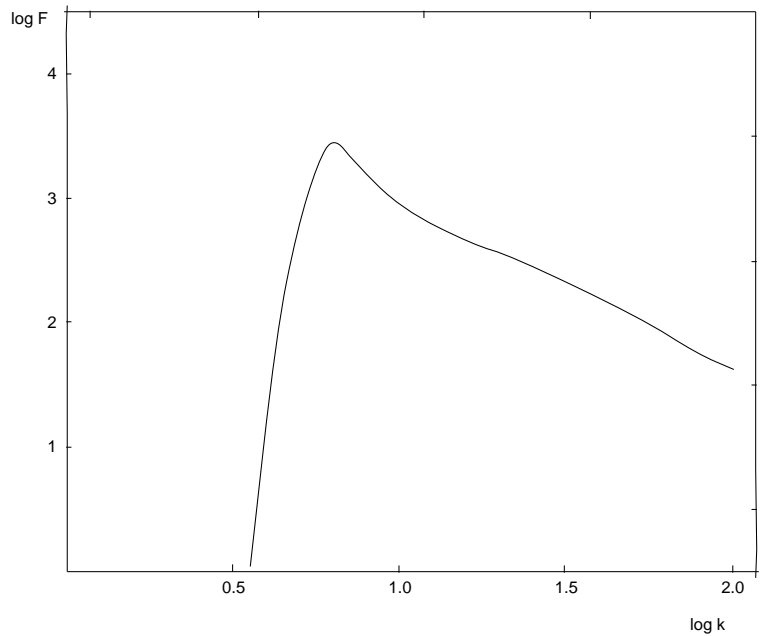,width=10cm,angle=270}}   	
   \caption{The $F(k)$ angle-integrated double energy spectral function 
for the case $S=10^5$, $k_{co}=3.14$, $Q=1$.  Note the backscatter at low 
wavenumbers.}
\end{figure}

In Figure 6 the
\[ F(k)= 2\pi\int_0^\pi F(k,\theta)k^2 \sin\theta\,d\theta   \]
integrated double energy spectral function is shown.  This figure is to be 
compared with the Figure 23 of the CGC paper.  Contrary to the isotropic GISS 
model, some backscatter is noticeable in the present model, though it is not so 
emphasized as in the DIA.

\begin{figure}
   \centerline{\psfig{figure=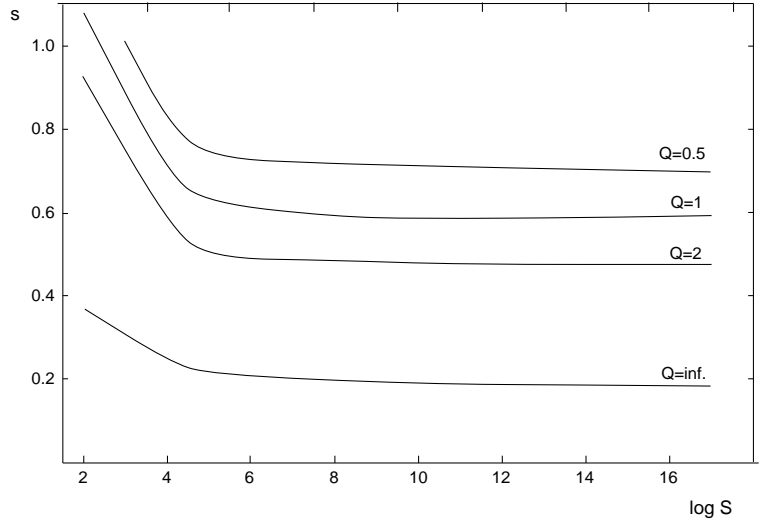,width=10cm,angle=270}}   	
   \caption{The $s$ anisotropy parameter as a function of $S$ for 
different $Q$ values.}
\end{figure}

Values of the $s$ anisotropy parameter for different models are shown in Figure 
7 as a function of $S$.  These values can be computed from
\[ 2s+1=\frac{\int_0^{k_{max}}\,dk \int_0^\pi \alpha(k, 
   \theta, \phi) F(k, \theta, \phi) \sin\theta\,d\theta}{\int_0^{k_{max}}\,dk 
   \int_0^\pi F(k, \theta, \phi) \sin^3\theta\,d\theta}.
    \]
i.\  e.\  the 2D version of Equation (34).  It is apparent that instead of going 
to zero, $s$ saturates to 
a finite limiting value as $S\rightarrow\infty$.  This qualitative behavior is 
independent of the choice of the $Q$ quenching parameter, though the actual 
asymptotic value of $s$ does depend on $Q$.  The value $s=0.37$ found in Chan 
and Sofia's (1989) simulations corresponds to $Q\sim 3$--$4$ but the anisotropy 
remains quite moderate even with $Q\rightarrow \infty $.  A similar effect, with 
a weaker dependence on $Q$ is observed in the shape of the $x(S)$ function 
(Figure 8).  The $x$ quantity defined in (1) was calculated here using
\begin{eqnarray}
   &&l_x^{-1}= 2\pi\int_0^{k_{max}} dk \int_0^\pi F(\vc k) k_x k^2 
   \sin\theta\,d\theta \nonumber \\   
   &&l_z^{-1}= 2\pi\int_0^{k_{max}} dk \int_0^\pi F(\vc k) k_z k^2 
   \sin\theta\,d\theta .   
\end{eqnarray}

\begin{figure}
   \centerline{\psfig{figure=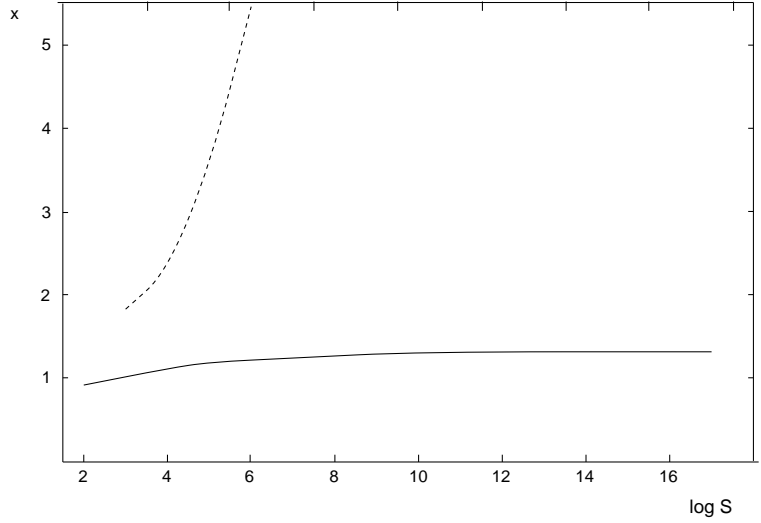,width=10cm,angle=270}}   	
   \caption{The $x$ anisotropy parameter as a function of $S$ in the 
present model (continuous line) and for the most unstable linear mode on the 
basis of the formulas of Canuto (1990) (dashed line).}
\end{figure}

\begin{figure}
   \centerline{\psfig{figure=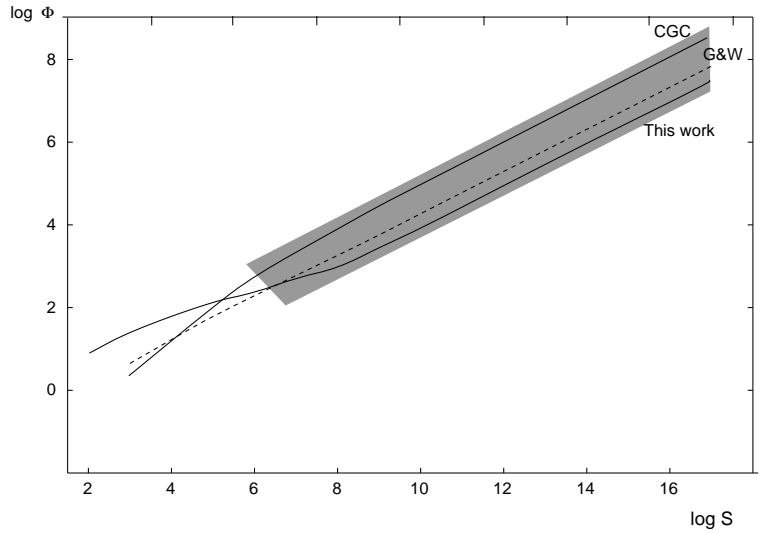,width=10cm,angle=270}}   	
   \caption{The $\Phi$ nondimensional convective flux as a function of $S$ 
for the present work, for the CGC model, for the mlt formula of Gough and Weiss 
(1976), and the observational limits deduced by Spruit (1974) (dashed area). The
mlt fluxes were nondimensionalized assuming 
$l=H_P$.}
\end{figure}

In Figure 9 the $S$-dependence of the nondimensional convective flux 
\[ \Phi= \frac{2\pi}S \int_0^\infty dk \int_0^\pi n_S(\vc k)\alpha(\vc k)
   F(\vc k) k^2\sin\theta\,d\theta      \]
is shown. For comparison, the CGC result and some  
mixing-length theory (mlt) calibrations (with the choice $l=H_P=D$ as length 
scale) are also shown.  It should be noted though that a direct comparison of 
the present Boussinesq model with mlt models of the strongly stratified solar 
convective zone is not necessarily sensible.
The computed curve lies near the lower limit 
derived by Spruit (1974) and is by a factor of 7 lower (at large $S$ values)
than the isotropic CGC curve.  Similarly, the total specific kinetic energy
\[ K= \pi \int_0^{k_{max}} dk \int_0^\pi F(\vc k) k^2 \sin\theta\,d\theta
    \]
(Figure 10) and the turbulent viscosity
\[ \nu_t= 2\pi\int_0^{k_{max}} dk  \int_0^\pi 
   \frac{F(\vc k) k^2}{n_c(\vc k)} \sin\theta\,d\theta    \]
(Figure 11) are also considerably lower than the isotropic results.
From these results the coefficient of the $K$--$\epsilon$ relation
\[ \epsilon= \xi_3 K^2/\nu_t      \]
is found to be $\xi_3=0.184$, which is a factor of two higher than the 
experimental value.  Similarly, for Smagorinsky's constant we get $C\simeq 
0.001$, to be compared with the latest experimental value of 0.003 (quoted by 
CGC).  As in CGC, the agreement of the computed coefficients with the 
experimental ones to within a factor of 2 or 3 indicates that the qualitative 
behavior of $n_S(\vc k)$  is typical for a general class of mechanisms 
generating turbulence.

\begin{figure}
   \centerline{\psfig{figure=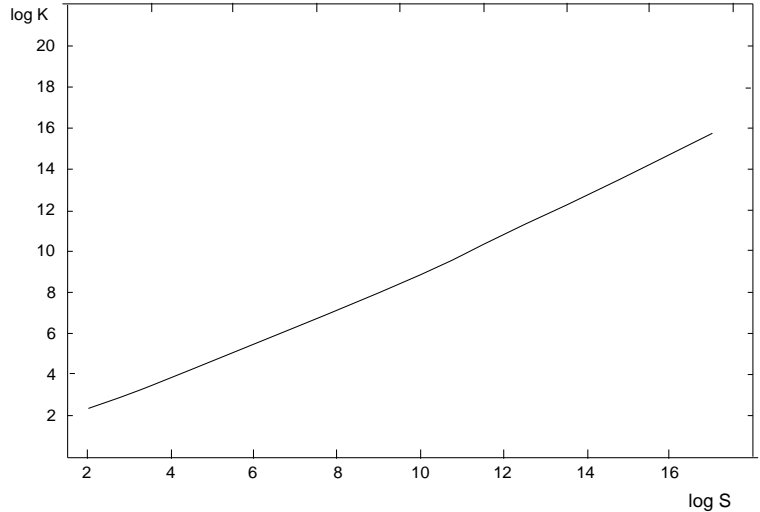,width=10cm,angle=270}}   	
   \caption{The $K$ specific turbulent kinetic energy as a function of 
$S$.}
\end{figure}

\begin{figure}
   \centerline{\psfig{figure=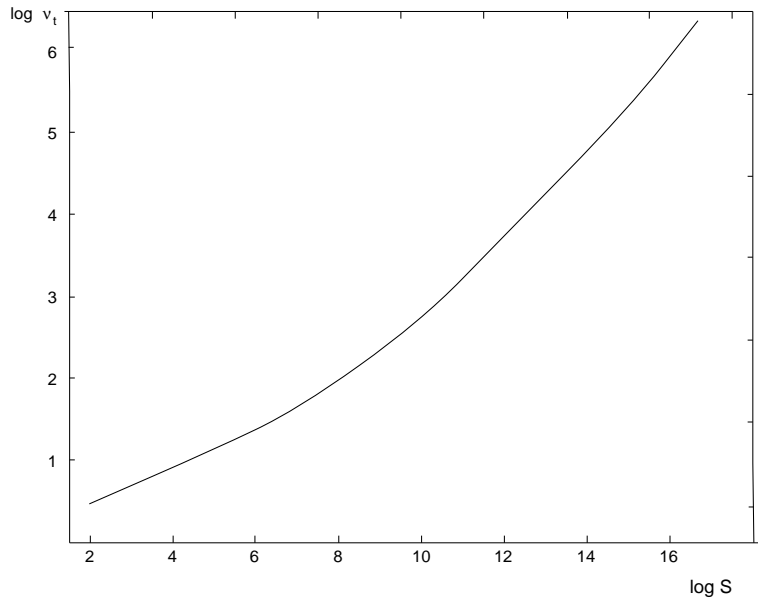,width=10cm,angle=270}}   	
   \caption{The $\nu_t$ turbulent viscosity as a function of $S$.}
\end{figure}

\section{CONCLUSION}
The application of the homogeneous anisotropic incompressible turbulence model 
desribed in Section 2  has shown that the anisotropy of turbulent thermal 
convection does \it not\/ \rm grow indefinitely with incresing $S$, but it 
rather saturates to a finite (and moderate) limiting value.  This result is in 
agreement with 
numerical experiments but it is in contradiction with earlier theoretical 
expectations based on linear stability analysis.  It remains to be seen how the 
inhomogeneity (nonlocality) of turbulence will influence these results, but it 
appears that the assumption of near isotropy is not the main factor limiting 
the validity of present-day convection theories.

The computations show that the resulting $\Phi$ convective flux is practically 
insensitive to the value of the $Q$ quenching parameter, unless $Q\ll 1$.  The 
physical reason for this is that the flux is predominantly determined by the 
low wavenumber energy-containing eddies, so details of the quenching at higher 
wavenumbers do not have serious effects on it.  The comparison with numerical 
experiments shows that $Q$ is indeed of order unity.  The low sensitivity of 
$\Phi$ to $Q$ means that the $\Phi'opto S^{1/2}$ scaling remains valid for 
any form of the (unknown) $Q(S)$ function.  As the $Q$ parameter basically 
determines the efficiency of the coupling between different velocity components 
(the ``toroidal'' and ``poloidal'' components of Massaguer, 1991), it follows 
that the secondary instabilities generating toroidal motions do not 
significantly influence the energy transport.

The work reported here has been the first attempt to study anisotropic 
turbulent convection  in a self-consistent (though approximate) way.  It is 
needless to say that further, more sophisticated models are necessary to extend 
these studies, in particular to the inhomogeneous case.

\bigskip

\noindent
\it Acknowledgement

\rm\bigskip\nopagebreak
\noindent
I am grateful for Carole Jordan's encouragement.  This work was partly funded 
by a Soros/FCO grant and by the OTKA grant No. 2135.

{}

\end{document}